\documentclass[aip,jcp,reprint,groupedaddress]{revtex4-1}

\pdfoutput=1

 \usepackage{amsmath}
 \usepackage{amsfonts}
 \usepackage{graphicx}
 \usepackage{newtxtext,newtxmath}
 \usepackage{comment}
\usepackage{dcolumn}
\usepackage{natbib}

 \usepackage{hyperref}
 \hypersetup{
    colorlinks=true,       
    linkcolor=blue,          
    citecolor=blue,        
    filecolor=magenta,      
    urlcolor=blue           
}

\newcommand{\red}[1]{{\textcolor{red}{#1}}}

\renewcommand{\vec}[1]{\boldsymbol{\mathbf{#1}}}

\renewcommand{\epsilon}{\varepsilon}

\newcolumntype{d}[1]{D{.}{.}{#1}}

\let\originalleft\left
\let\originalright\right
\renewcommand{\left}{\mathopen{}\mathclose\bgroup\originalleft}
\renewcommand{\right}{\aftergroup\egroup\originalright}

\renewcommand{\epsilon}{\varepsilon}
\renewcommand{\vec}[1]{\boldsymbol{\mathbf{#1}}}

\begin{document}

\frenchspacing

\title{Calculations of positron binding and annihilation in polyatomic molecules}
\author{A.~R. Swann}
\email{a.swann@qub.ac.uk}
\author{G.~F. Gribakin}
\email{g.gribakin@qub.ac.uk}
\affiliation{
School of Mathematics and Physics, Queen's University Belfast, University Road, Belfast BT7 1NN, United Kingdom}
\date{\today}

\begin{abstract}
A model-potential approach to calculating positron-molecule binding energies and annihilation rates is developed. Unlike existing \textit{ab initio} calculations, which have mostly been applied to strongly polar molecules, the present methodology can be applied to both strongly polar and weakly polar or nonpolar systems. The electrostatic potential of the molecule is calculated at the Hartree-Fock level, and a model potential that describes short-range correlations and long-range polarization of the electron cloud by the positron is then added. The Schr\"odinger equation for a positron moving in this effective potential is solved to obtain the binding energy. The model potential contains a single adjustable parameter for each type of atom present in the molecule. The wave function of the positron bound state may be used to compute the rate of electron-positron annihilation from the bound state. As a first application, we investigate positron binding and annihilation for the hydrogen cyanide (HCN) molecule. Results for the binding energy are found to be in accord with existing calculations, and we predict the rate of annihilation from the bound state to be $\Gamma=0.1\text{--}0.2 \times 10^9~\text{s}^{-1}$.
\end{abstract}

\maketitle

\section{\label{sec:intro}Introduction}

The aim of this paper is to develop an approach that would enable reliable calculations of positron bound states with polyatomic molecules.

Since  the positron ($e^+$) was predicted in 1931\cite{Dirac31} and discovered in 1933,\cite{Anderson33} it has proved to be a useful tool in many areas of science, including fundamental tests of QED and the standard model,\cite{Karshenboim05,Ishida14,ALEPH06} astrophysics,\cite{Guessoum14} condensed-matter physics,\cite{Tuomisto13} atomic physics,\cite{Charlton01} and medicine.\cite{Wahl02} The physics and chemistry of positron and positronium (Ps, an electron-positron bound pair) has seen much progress from the important early advances\cite{Goldanskii68} to many new directions envisaged at the turn of the century.\cite{Antimatter01} Despite this, there is much about positron interactions with ordinary matter that is still not well understood. One open problem is positron binding to atoms and molecules. 

The possibility of positron binding to neutral atoms was predicted by many-body-theory calculations in 1995.\cite{Dzuba95} This was subsequently confirmed by variational calculations of the $e^+\mathrm{Li}$ system,\cite{Ryzhikh97,Strasburger98} and calculations of positron binding to other atoms soon appeared.\cite{Mitroy02,Dzuba12,Harabati14} However, no experimental evidence of positron-atom bound states has yet arisen. Several methods of producing such states have been proposed,\cite{Mitroy99,Dzuba10,Surko12,Swann16} but difficulties regarding the limited availability of suitable positron sources, the need to obtain the neutral-atom species in the gas phase, and implementation of an unambiguous detection scheme have so far prevented detection.

The situation for positron binding to molecules is radically different. Positron-molecule binding energies can be measured by virtue of the process of \textit{resonant annihilation}. When a positron collides with a polyatomic molecule, two annihilation mechanisms are possible: direct, ``in flight'' annihilation of the positron with one of the target electrons, and resonant annihilation, where the positron is captured into a quasibound state, with any excess energy being transferred into molecular vibrations, typically those of a mode with near-resonant energy.\cite{Gribakin00,Gribakin01,Gribakin10}  Resonant annihilation is operational for molecules that are capable of binding the positron. It leads to pronounced peaks in the positron-energy dependence of the annihilation rate.\cite{Gilbert02} Observation of resonances with energies
\begin{equation}
\epsilon_\nu = \hbar\omega_\nu - \epsilon_b ,
\end{equation}
where $\omega_\nu$ is the frequency of vibrational mode $\nu$,
has enabled measurement of the positron binding energies $\epsilon_b$ for over 70 molecules.\cite{Barnes03,Barnes06,Young08,Young08a,Danielson09,Danielson10,Danielson12,Natisin:thesis} The majority of these are nonpolar or weakly polar species, such as alkanes and related hydrocarbons, aromatics, partially halogenated hydrocarbons, alcohols, formates, and acetates. Analysis of the experimental data obtained prior to 2009 led to the following empirical formula for the positron binding energy:
\begin{equation}\label{eq:bin_en_empirical}
\epsilon_b = 12.4 (\alpha + 1.6\mu - 5.6),
\end{equation}
where $\alpha$ is the dipole polarizability of the molecule in units of $10^{-24}$~cm$^3$, $\mu$ is the dipole moment of the molecule in debyes (D), and  $\epsilon_b$ is in units of meV.\cite{Danielson09} More recent data  have highlighted the deficiency of this fit, with Eq.~(\ref{eq:bin_en_empirical}) underestimating the binding energies for strongly polar molecules.\cite{Danielson12}

On the side of theory, calculations of positron-molecule binding energies have proven to be very challenging. It is known that a static molecule with dipole moment $\mu>1.625$~D possesses an infinite number of positron (as well as electron) bound states.\cite{Fermi47,Crawford67} (For a molecule that is free to rotate, the critical value of the dipole moment increases with the angular momentum of the molecule.\cite{Garrett71}) This means that positron binding to strongly polar molecules is obtained even at the lowest, static-potential level of the theory. However, prior to the experimental observation of resonant annihilation, there were few attempts at this problem. Predictions of binding were made for strongly polar molecules using semiempirical\cite{SW76} and Hartree-Fock (HF)\cite{Kurtz81,Tachikawa01} methods. The effect of correlations on the feasibility of binding was explored using the $R$-matrix method,\cite{DT88} configuration interaction (CI),\cite{Strasburger96} explicitly correlated Gaussian functions (ECG),\cite{Strasburger99} and quantum Monte Carlo (QMC)\cite{Bressanini98,Mella01} (for LiH and a few other polar diatomics and H$_2$O).

By contrast, from 2002 onwards, many papers on positron binding to molecules have been published by several quantum-chemistry groups. The majority of the calculations are for simple diatomic and triatomic molecules, e.g., alkali hydrides,\cite{MMBE00,S01,BA04,BLMTPK05,GFBLPRTK06,KMTTN11} metal oxides,\cite{Bressanini98,BLPTK07,BL08} HCN,\cite{Chojnacki06,Kita09} CXY (X, Y = O, S, Se),\cite{Koyanagi13} and formaldehyde.\cite{S04,Tachikawa12,YKT14} However, a number of calculations also examined binding to larger species, such as urea and acetone,\cite{Tachikawa03}  nitriles,\cite{Tachikawa11} and aldehydes.\cite{Tachikawa12} There are also exploratory studies for amino acids \cite{KKT12,CRVR14} and nucleic bases and pairs,\cite{KKST13,RCFVR14} some of which apply the any-particle--molecular-orbital (APMO) framework to include correlation effects using a many-body theory approach.\cite{CRVR14,RCFVR14} In particular, these calculations showed that the binding energies obtained at the static HF level increase considerably when electron-positron correlations are included, e.g., for acetonitrile CH$_3$CN, $\epsilon_b$ increases from 15~meV (HF) to 135~meV (CI).\cite{Tachikawa11} 

In spite of the large number of calculations, at present, only six molecules, namely, carbon disulfide CS$_2$, acetaldehyde C$_2$H$_4$O, propanal C$_2$H$_5$CHO, acetone (CH$_3$)$_2$CO, acetonitrile CH$_3$CN, and propionitrile C$_2$H$_5$CN, have been studied both experimentally\cite{Danielson10,Danielson12} and theoretically.\cite{Tachikawa03,Tachikawa11,Tachikawa12,Koyanagi13,Tachikawa14} The closest agreement between theory and experiment is for acetonitrile, whose measured binding energy is 180~meV,\cite{Danielson10} some 33\% larger than the CI result.\cite{Tachikawa11} The biggest discrepancy is for carbon disulfide, the only nonpolar molecule on this list, where the measured binding energy is 75~meV,\cite{Danielson10} while the calculations predict no binding.\cite{Koyanagi13} These discrepancies show the great difficulty in providing an accurate description of the electron-positron correlations, especially for nonpolar molecules, where there is no binding at the lowest (static) level of theory.

As far as we are aware, there are no successful \textit{ab initio} calculations of positron binding to weakly polar ($\mu<1.625$~D) or nonpolar molecules, where binding has been seen in experiment, and where it is enabled exclusively by electron-positron correlation effects. Gribakin and Lee modeled positron binding to the $n$-alkanes (C$_n$H$_{2n+2}$) using a zero-range-potential (ZRP) approach.\cite{Gribakin06} By fitting the ZRP parameter to reproduce the measured binding energy for dodecane ($n=12$),\cite{Gribakin09} they obtained a good overall description of the problem. However, some quantitative details were not captured correctly: binding was predicted for $n\geq4$, with a second bound state emerging for $n\geq13$, while experimentally, binding is measured for $n=3$ already, with a second bound state for $n=12$.\cite{Barnes06,Young08}

In this work, a model-potential method is developed to calculate positron-molecule binding energies. First, the electrostatic potential of the molecule is calculated at the HF level. The Schr\"odinger equation is then solved for a positron moving in this potential, with the addition of a model potential that accounts for the long-range polarization of the molecule and short-range correlations. The method can be applied to both strongly polar molecules and weakly polar or nonpolar molecules.
While this is not an \textit{ab initio} technique, it broadly captures the essential physics of the positron-molecule interaction and enables calculations to be carried out with much less computational expense than \textit{ab initio} methods.
A similar approach has previously been shown to accurately describe positron scattering, annihilation, and (when it exists) binding  in noble-gas and other closed-shell atoms.\cite{Mitroy02a} As a first application, we consider positron binding to hydrogen cyanide HCN and make comparisons with existing calculations. We also use the positron wave function to calculate the rate of annihilation from the bound state.

Except where otherwise stated, atomic units (a.u.) are used throughout; the atomic unit of length (the Bohr radius) is denoted by $a_0$.

\section{Theory}

\subsection{Hartree-Fock methods}

The nonrelativistic Hamiltonian for a positron interacting with a molecule consisting of $N_e$ electrons and $N_a$ nuclei (treated in the Born-Oppenheimer approximation) is
\begin{align}\label{eq:e+mol_exact_hamiltonian}
H &= \sum_{i=1}^{N_e} h^e(\vec{r}_i) + h^p(\vec{r}) + \sum_{i=1}^{N_e} \sum_{j<i} \frac{1}{\lvert \vec{r}_i - \vec{r}_j \rvert}
- \sum_{i=1}^{N_e} \frac{1}{\lvert \vec{r} - \vec{r}_i \rvert} ,
\end{align}
where
\begin{align}
h^e(\vec{r}_i) &= -\frac12 \nabla_i^2 - \sum_{A=1}^{N_a} \frac{Z_A}{\lvert \vec{r}_i - \vec{r}_A \rvert} , \\
h^p(\vec{r}) &= -\frac12 \nabla_p^2  + \sum_{A=1}^{N_a} \frac{Z_A}{\lvert \vec{r} - \vec{r}_A \rvert} , \label{eq:hp}
\end{align}
$\vec{r}_i$ is the position of electron $i$, $\vec{r}_A$ is the position of nucleus $A$ (with charge $Z_A$), and $\vec{r}$ is the position of the positron, all relative to an arbitrary origin. A direct solution of the Schr\"odinger equation,
\begin{equation}
H \Psi(\vec{r}_1,\vec{r}_2,\dotsc,\vec{r}_{N_e},\vec{r})
=E \Psi(\vec{r}_1,\vec{r}_2,\dotsc,\vec{r}_{N_e},\vec{r}) ,
\end{equation}
for the system energy $E$ and wave function $\Psi$ is prevented by the electron-electron and electron-positron Coulomb interactions [the final two terms in Eq.~(\ref{eq:e+mol_exact_hamiltonian})] that make this numerically intractable for systems with more than a few electrons.

The starting point for our calculations of positron-molecule binding is the HF method. We assume that the molecule is closed-shell; thence there are $N_e/2$ doubly occupied molecular orbitals $\varphi_i(\vec{r}_i)$. We consider two distinct ways in which the HF method can be applied.

\textit{Frozen-target method.}---In this case, the energy and wave function of the \textit{bare} molecule (i.e., without the positron) in the ground state are computed in the conventional HF approximation. This wave function $\Phi(\vec{r}_1,\vec{r}_2,\dotsc,\vec{r}_{N_e})$ is a Slater determinant of the $N_e$ spin orbitals. The Schr\"odinger equation for a positron moving in the resulting electrostatic potential of the molecule is then 
\begin{equation}\label{eq:electrostatic_pot}
\left[ h^p(\vec{r})
- 2\sum_{i=1}^{N_e/2} J_i^e(\vec{r}) \right] \psi(\vec{r}) = \epsilon_p \psi(\vec{r}),
\end{equation}
where
\begin{equation}
J_i^e(\vec{r}) = \int \frac{\lvert \varphi_i(\vec{r}') \rvert^2}{\lvert \vec{r} - \vec{r}' \rvert} \, d^3\vec{r}'.
\end{equation}
This
is solved to find the positron energy $\epsilon_p$ and wave function $\psi(\vec{r})$. The total wave function of the system is given by
\begin{equation}\label{eq:system_wave_function_product}
\Psi(\vec{r}_1,\vec{r}_2,\dotsc,\vec{r}_{N_e},\vec{r}) = \Phi(\vec{r}_1,\vec{r}_2,\dotsc,\vec{r}_{N_e}) \psi(\vec{r}) .
\end{equation}
The key feature of this approach is that the electrons are ``unaware'' of the presence of the positron. That is, the electronic molecular orbitals are calculated in the static mean-field approximation, and distortion of the electronic molecular orbitals by the positron is not accounted for at all. We refer to this as the \textit{frozen-target} (FT) method.

\textit{Relaxed-target method.}---Here the wave function of the system is again assumed to take the form of Eq.~(\ref{eq:system_wave_function_product}). A modified version of the HF method that accounts for the presence of the positron is used to compute the electron wave functions. The modified HF equations for the electrons are
\begin{equation}\label{eq:MHF_ele}
\left\{ h^e(\vec{r}_i)
+ \sum_{\substack{j=1\\ j\neq i}}^{N_e/2} [2J_j^e(\vec{r}_i) - K_j^e(\vec{r}_i)]
- J^p(\vec{r}_i) \right\} \varphi_i(\vec{r}_i) = \epsilon_i \varphi_i(\vec{r}_i),
\end{equation}
where $i=1,2,\dotsc,N_e/2$,
\begin{align}
K_j^e(\vec{r}_i)\varphi_i(\vec{r}_i) &= \varphi_j(\vec{r}_i) \int \frac{\varphi_j^*(\vec{r}') \varphi_i(\vec{r}')}{\lvert \vec{r}_i - \vec{r}' \rvert} \, d^3\vec{r}' , \\
J^p(\vec{r}_i) &= \int \frac{\lvert \psi(\vec{r}') \rvert^2}{\lvert \vec{r}_i - \vec{r}' \rvert} \, d^3\vec{r}' .
\end{align}
The corresponding   equation for the positron is identical to Eq.~(\ref{eq:electrostatic_pot}).
It is clear from Eqs.~(\ref{eq:MHF_ele}) and (\ref{eq:electrostatic_pot}) that the motions of the electrons and the positron are coupled: the positron density appears in the modified HF equations for the electrons, and vice versa.
Equations (\ref{eq:MHF_ele}) and (\ref{eq:electrostatic_pot}) are solved self-consistently and simultaneously to obtain the $\varphi_i(\vec{r}_i)$ and $\psi(\vec{r})$.
This approach is the foundation of CI calculations of positron-molecule binding. It has also been  used in explicitly correlated HF studies of the 
PsH,\cite{Pak09,Swalina12} LiPs,\cite{Swalina12,Sirjoosingh13} and $e^+$LiH\cite{Swalina12,Sirjoosingh13} systems.
To contrast with the FT method,  the electronic molecular orbitals are now ``aware'' of the presence of the positron, but the electron-positron interaction is still only treated at the static, mean-field level: the dynamical electron-positron correlations (which are responsible for long-range polarization of the molecule by the positron) are still not accounted for.
We refer to this as the \emph{relaxed-target} (RT) method.

The positron binding energy $\epsilon_b$ in either method is given by the difference between the energy of the bare molecule $M$ and the energy of the bound positron-molecule system $e^+M$, viz.,
\begin{equation}
\epsilon_b = E(M) - E(e^+M).
\end{equation}
In the FT method, this is equal to the negative of the energy of the  positron orbital, i.e.,
\begin{equation}
\epsilon_b = - \epsilon_p.
\end{equation}
Note that both methods are approximations: dynamical electron-electron and electron-positron correlations have been neglected. Consequently, only molecules with dipole moments greater than $1.625$~D can bind a positron at this level of approximation. The RT approximation will always give a slightly larger value of $\epsilon_b$ than the FT approximation, since the molecular electron cloud has the freedom to distort such that the total energy of the system is minimized.

\subsection{Model correlation potential}

As was stated in Sec.~\ref{sec:intro}, failure to account for the dynamical electron-electron and electron-positron correlations leads to a lack of binding for weakly polar molecules and seriously underestimated values of $\epsilon_b$ even for strongly polar molecules. Physically, the interaction between the positron and the molecule can be cast as the sum of two terms, viz.,
\begin{equation}\label{eq:e+_twoterm_int}
V(\vec{r}) = V_\text{st}(\vec{r}) + V_\text{cor}(\vec{r}),
\end{equation}
where [see Eqs.~(\ref{eq:hp}) and (\ref{eq:electrostatic_pot})]
\begin{equation}\label{eq:Vst}
V_\text{st}(\vec{r}) = \sum_{A=1}^{N_a} \frac{Z_A}{\lvert \vec{r} - \vec{r}_A \rvert} - 2\sum_{i=1}^{N_e/2} J_i^e(\vec{r}) 
\end{equation}
is the static potential of the molecule,
and $V_\text{cor}(\vec{r})$ accounts for the residual interactions absent in the HF methods. The exact form of $V_\text{cor}(\vec{r})$ (which can be derived using many-body theory\cite{Dzuba95,Dzuba96,Gribakin04,Green14}) is very difficult to compute exactly.\footnote{The true correlation potential is a nonlocal and energy-dependent operator, see, e.g., Figs. 2 and 3 in Ref. \onlinecite{Green18a}.} However, at distances far from the molecule it takes the simple asymptotic form
\begin{equation}
V_\text{cor}(\vec{r}) \simeq -\frac{1}{2r^6} \sum_{i,j}x_ix_j \alpha_{ij} ,
\end{equation}
where the $x_i$ ($i=1$, 2, 3) are the Cartesian coordinates $x$, $y$, and $z$ of the positron as measured from the molecule, the $\alpha_{ij}$ are the Cartesian components of the molecule's dipole polarizability tensor, and $r=(x^2+y^2+z^2)^{1/2}$. This describes polarization of the molecule by the positron. For spherically symmetric targets (e.g., closed-shell atoms) and spherical-top molecules, the polarizability tensor is isotropic, and
\begin{equation}
V_\text{cor}(\vec{r}) \simeq - \frac{\alpha}{2r^4} ,
\end{equation}
where $\alpha$ is the scalar dipole polarizability.

Calculations for noble-gas and other closed-shell atoms  show that positron scattering, annihilation, and binding can be successfully described by using a model correlation potential of the form\cite{Mitroy02a}
\begin{equation}\label{eq:model_polarization_potential}
V_\text{cor}(\vec{r}) = -\frac{\alpha}{2r^4} \left[ 1-\exp\bigl( -r^6/\rho^6 \bigr) \right] .
\end{equation}
The function in brackets moderates the unphysical growth of the potential at small $r$, with $\rho$ a cutoff parameter whose values are fitted to reproduce the results of more sophisticated scattering or bound-state calculations. The short-range part of $V_\text{cor}(\vec{r})$ allows one to account for other correlation effects, such as virtual positronium formation. Values of $\rho$ correlate with the radius of the atom, e.g., $\rho =1.50a_0$ for He, $2.05a_0$ for H, and $3.03a_0$ for Mg.\cite{Mitroy02a}

In this work, we construct a model positron-molecule correlation potential as a sum of potentials of the form of Eq.~(\ref{eq:model_polarization_potential}), centered on each of the molecule's constituent atoms, viz.,
\begin{equation}\label{eq:molecular_corr_potential}
V_\text{cor}(\vec{r}) = -\sum_{A=1}^{N_a} \frac{\alpha_A}{2 \lvert \vec{r}-\vec{r}_A\rvert^4}
\left[ 1-\exp\left( -\frac{\lvert \vec{r}-\vec{r}_A\rvert^6}{\rho_A^6}\right)\right],
\end{equation}
where $\alpha_A$ is the \textit{hybrid} polarizability\cite{Miller90} of atom $A$ within the molecule, and $\rho_A$ is a cutoff radius specific to atom $A$. The atomic hybrid polarizabilities $\alpha_A$ take into account the chemical environment of the atom in a molecule, and their sum $\alpha =\sum _A\alpha _A$ yields the total polarizability of the molecule.

A natural and important question in this approach to the positron-molecule binding problem is whether $V_\text{st}(\vec{r})$, which appears in Eq.~(\ref{eq:e+_twoterm_int}), should be computed using the FT approximation or the RT approximation. The model correlation potential (\ref{eq:molecular_corr_potential}) is designed to account for the dynamical distortion of the electron cloud by the positron in an approximate way. Therefore, if $V_\text{st}(\vec{r})$ is calculated using the RT method (where limited distortion of the electron cloud by the positron is already included at the HF level), there will be an effective overestimation of the correlation effects. Thus we use the model correlation potential in conjunction with $V_\text{st}(\vec{r})$ as found using the FT method.

In practice, this is a two-step process. First, the electronic orbitals of the bare molecule (i.e., without the positron) are computed using the conventional HF method. Then, the Schr\"odinger equation for the positron,
\begin{equation}
\left[ h^p(\vec{r}) -2 \sum_{i=1}^{N_e/2} J^e_i(\vec{r})+ V_\text{cor}(\vec{r}) \right] \psi(\vec{r}) = \epsilon_p \psi(\vec{r}) ,
\end{equation}
is solved to obtain the energy and wave function of the positron bound state. We hereafter refer to this as the \textit{frozen-target-plus-polarization} (FT+P) approximation. Note that this is consistent with the many-body theory approach which starts with the HF calculation of the target in the ground state. Its potential is then used to generate sets of excited electron and positron states for the subsequent calculation of the correlation potential and positron (Dyson) wave function.\cite{Green14}

\subsection{Annihilation rate}

The  $2\gamma$ annihilation rate for a positron bound to a molecule (or atom) with the zero electron spin is given by\cite{Gribakin10}
\begin{equation}\label{eq:Gamma_from_delta}
\Gamma = \pi r_0^2 c \delta_{ep},
\end{equation}
where $r_0$ is the classical electron radius, $c$ is the speed of light, and $\delta_{ep}$ is the average electron density at the position of the positron:
\begin{align}\label{eq:con_dens_exact}
\delta_{ep} = \int \sum_{i=1}^{N_e} \delta(\vec{r}-\vec{r}_i) \lvert \Psi(\vec{r}_1,\dotsc,\vec{r}_{N_e},\vec{r}) \rvert^2\, d^3\vec{r} \prod _{j=1}^{N_e} d^3\vec{r}_j.
\end{align}
Here, $ \Psi(\vec{r}_1,\dotsc,\vec{r}_{N_e},\vec{r})$ is the total wave function for the $N_e$ electrons and the positron, normalized as 
\begin{equation}
\int \lvert \Psi(\vec{r}_1,\dotsc,\vec{r}_{N_e},\vec{r}) \rvert^2\, d^3\vec{r}\prod _{j=1}^{N_e} d^3\vec{r}_j = 1.
\end{equation}
The contact density has units of inverse volume, so it is expressed in terms of $a_0^{-3}$ when atomic units are in use.

For the wave function in the form of Eq.~(\ref{eq:system_wave_function_product}) (sometimes referred to as the independent-particle approximation), Eq.~(\ref{eq:con_dens_exact}) becomes
\begin{equation}\label{eq:con_den_IPA}
\delta_{ep} = 2\sum_{i=1}^{N_e/2} \int \lvert \varphi_i(\vec{r}) \rvert^2 \lvert \psi(\vec{r})\rvert^2 \, d^3\vec{r}.
\end{equation}
The annihilation rate can thus be straightforwardly calculated from the wave functions of the molecular orbitals and the bound positron state. However, the independent-particle approximation does not account for short-range correlations that increase the density of the electrons at the positron, and consequently Eq.~(\ref{eq:con_den_IPA}) underestimates the true value of $\delta_{ep}$.\footnote{In the many-body-theory approach such correlations are represented by the annihilation-vertex corrections.\cite{Gribakin04,Green14,Green15,Dunlop06}} This shortcoming can be alleviated by introducing molecular-orbital-specific \textit{enhancement factors} $\gamma_i$ into Eq.~(\ref{eq:con_den_IPA}), viz.,
\begin{equation}\label{eq:con_den_enh}
\delta_{ep} = 2\sum_{i=1}^{N_e/2} \gamma_i \int \lvert \varphi_i(\vec{r}) \rvert^2 \lvert \psi(\vec{r})\rvert^2 \, d^3\vec{r},
\end{equation}
where $\gamma_i \geq 1$. Similar enhancement factors are used in calculations of positron annihilation in solids.\cite{Puska94,Alatalo96}

Green and Gribakin\cite{Green15,Green18} used many-body perturbation theory to calculate  enhancement factors for positron annihilation in noble-gas atoms. These enhancement factors were computed for positive-energy positrons and were found to be approximately constant for energies $\lesssim$1~eV.\cite{Gribakin04} Their values were specific to the electron orbital and positron partial wave. In particular, it was found that the $s$-wave enhancement factors scale with the electron-orbital energy $\epsilon_i$ according to the empirical formula
\begin{equation}\label{eq:enh_fac_formula}
\gamma_i = 1 + \sqrt{\frac{1.31}{-\epsilon_i}} + \left( \frac{0.834}{-\epsilon_i} \right)^{2.15}.
\end{equation}
The positron bound to a polyatomic molecules does not have a well-defined orbital angular momentum. However, its wave function has a dominant $s$-wave character at small positron-atom separations, which provide the main contribution to the overlap intergals in Eq.~(\ref{eq:con_den_enh}). Hence, we shall use Eq.~(\ref{eq:enh_fac_formula}) to calculate the enhancement factors for annihilation in the positron-molecule bound state. 


\section{Numerical implementation}

The electron and positron wave functions are expanded in Gaussian basis sets centered on each of the atomic nuclei:
\begin{align}
\varphi_i(\vec{r}_i) &= \sum_{A=1}^{N_a} \sum_{k=1}^{N^e_{A}} C^{(i)}_{Ak} g_{Ak}(\vec{r}_i), \label{eq:ele_bas_exp}\\
\psi(\vec{r}) &= \sum_{A=1}^{N_a} \sum_{k=1}^{N^p_A} C^{(p)}_{Ak} g_{Ak}(\vec{r}), \label{eq:pos_bas_exp}
\end{align}
where
\begin{align}\label{eq:gaussian_def}
g_{Ak}(\vec{r}) = F_{Ak} (x-x_A)^{n^x_{Ak}} (y-y_A)^{n^y_{Ak}} (z-z_A)^{n^z_{Ak}}e^{-\zeta_{Ak} \lvert \vec{r} - \vec{r}_A\rvert^2}
\end{align}
is a Cartesian Gaussian basis function with angular momentum $n^x_{Ak} + n^y_{Ak} + n^z_{Ak}$ and normalization coefficient $F_{Ak}$, and there are $N_A^e$ ($N_A^p$) basis functions centered on each nucleus for the electron (positron).


For the electrons, we have used the standard 6-311++G($d$,$p$) basis set throughout. The equilibrium bond lengths are 1.059~\AA\ for H--C and 1.127~\AA\ for C${\equiv}$N.
For the positron, an even-tempered basis set is used:
\begin{equation}
\zeta_{Ak} = \zeta_{A1} \beta^{k-1} \qquad (k=1,\dotsc,N_A^p),
\end{equation}
where $\zeta_{A1}>0$ and $\beta>1$ are parameters (see Sec.~\ref{sec:FT_RT_results} for the values used). 
Correct choice of the smallest exponent $\zeta_{A1}$ for weakly bound positron states is very important. At large distances, the positron wave function behaves as $\psi (\vec{r})\propto e^{-\kappa r}$, where $\kappa =\sqrt{2\varepsilon _b}$. To ensure that expansion (\ref{eq:pos_bas_exp}) describes the wave function well at $r\sim 1/\kappa$, one must have $\zeta_{A1}\lesssim \kappa ^2 =2\varepsilon _b$.

The solution of the (modified) Roothaan equations for the electrons and positron is carried out in practice using \textsc{gamess}\cite{Schmidt93,Gordon05} with the \textsc{neo} package.\cite{Webb02,Adamson08} Modifications have been made to enable frozen-target calculations and to include the model correlation potential $V_\text{cor}(\vec{r})$ in the Roothaan equation for the positron.
To facilitate the computation of the matrix elements of the correlation potential, it is expressed as a sum of its constituent spherically symmetric atomic  potentials, viz.,
\begin{equation}\label{eq:pol_breakdown}
V_\text{cor}(\vec{r}) = \sum_{A=1}^{N_a} V_\text{cor}^{(A)}(|\vec{r}-\vec{r}_A|) ,
\end{equation}
where
\begin{equation}\label{eq:pol_pot_partial}
V_\text{cor}^{(A)}(r)  = - \frac{\alpha_A}{2 r^4}  \left[ 1-\exp\bigl( - r^6/\rho_A^6\bigr) \right].
\end{equation}
Each $V_\text{cor}^{(A)}(r)$ is  expanded in a set of $s$-type Gaussian functions:
\begin{equation}\label{eq:pol_pot_partial_expansion}
V_\text{cor}^{(A)}(r) = \sum_k D_{k}^{(A)} e^{-\kappa_{Ak} r^2},
\end{equation}
with the coefficients $D_{k}^{(A)}$ determined by a least-squares fit. For this, a set of 25 Gaussians has been used throughout, with exponents $\kappa_{Ak}=0.001$, 0.002, 0.004, 0.008, 0.016, 0.032, 0.064, 0.128, 0.256, 0.512, 1.0, 2.0, 3.0, 4.0, 5.0, 6.0, 7.0, 8.0, 9.0, 10.0, 20.0, 30.0, 40.0, 50.0, and 100.0.

\begin{figure}
\centering
\includegraphics{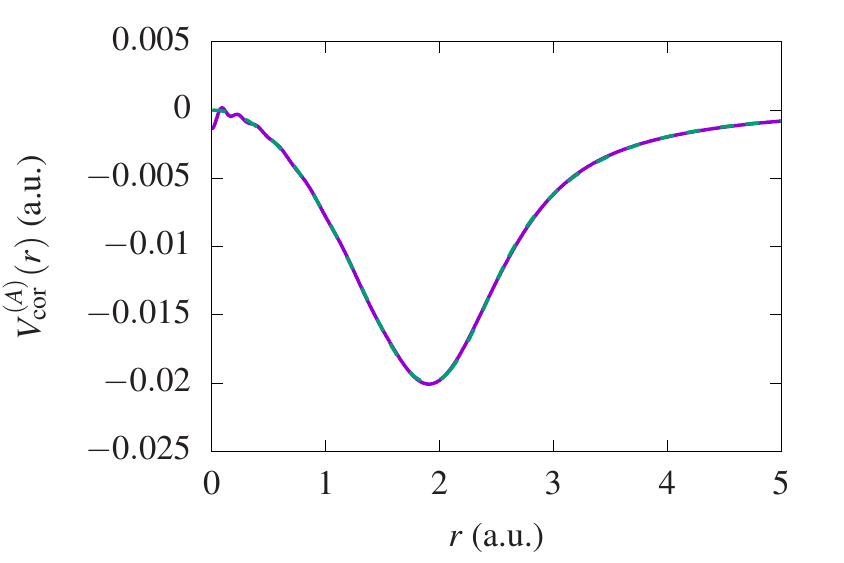}
\caption{\label{fig:pol_potential}Comparison of the the Gaussian-expanded form
of $V_\text{cor}^{(A)}(r)$ for $\alpha_A=1.0$~a.u. and $\rho_A=2.0$~a.u. [Eq.~(\ref{eq:pol_pot_partial_expansion}), solid purple curve] with its exact
analytical form [Eq.~(\ref{eq:pol_pot_partial}), dashed green curve].}
\end{figure}

Figure~\ref{fig:pol_potential} shows the analytical form (\ref{eq:pol_pot_partial}) of $V_\text{cor}^{(A)}(r)$ along with the Gaussian-expanded form (\ref{eq:pol_pot_partial_expansion}), for a polarizability of $\alpha_A=1.0$~a.u. and a (fairly typical) cutoff radius of $\rho_A=2.0$~a.u.
The two curves are indistinguishable on the scale of the graph, except at very small values of $r$, where the expansion (\ref{eq:pol_pot_partial_expansion}) exhibits some oscillations. These oscillations arise because the true form of $V_\text{cor}^{(A)}(r)$ goes to zero as $r\to 0$, while the $s$-type Gaussians in the expansion remain nonzero at $r=0$. The inclusion of Gaussians with large exponents (e.g., 50.0 and 100.0) is intended to give the expansion sufficient flexibility to approach zero as $r\to 0$, but the oscillations cannot be completely eradicated using a finite expansion. Note, however, that the positron wave function is strongly suppressed at small $\lvert \vec{r}-\vec{r}_A\rvert$, so that a small inaccuracy in the representation of $V_\text{cor}^{(A)}(r)$ near the origin has a negligible effect on the calculation of positron-molecule bound states.

Similarly, the expansion of $V_\text{cor}^{(A)}(r)$ in Gaussians cannot reproduce exactly the long-range asymptotic form of $V_\text{cor}^{(A)}(r)\simeq -\alpha_A/2r^4$. However, the inclusion of Gaussians with small exponents $\kappa_{Ak}$ in the expansion provides an accurate description of the long-range part. Indeed, a comparison of the value of $\int_0^\infty V_\text{cor}^{(A)}(r)\, dr$ for $\rho_A=2.0$~a.u., calculated using the analytical and Gaussian-expanded forms of $V_\text{cor}^{(A)}(r)$, reveals a difference of just 0.2\%.

Details of how matrix elements of $V_\text{cor}$ between positron basis functions and the electron-positron contact density are calculated are given in  Appendix~\ref{sec:app_a}.

\section{Results}

\subsection{\label{sec:FT_RT_results}FT and RT approximations}

As a test of this method, we investigate positron binding to hydrogen cyanide HCN. This molecule has a dipole moment of $2.98$~D\cite{CRC} and consequently can bind a positron even at the static level. Previous calculations of the $e^+$HCN binding energy have been carried out using the HF, CI, and QMC methods.\cite{Chojnacki06,Kita09} 

Table \ref{tab:HCN_FT_RT_bin_en} shows values of the positron binding energy $\epsilon_b$ in the FT and RT approximations. The positron basis set parameters used are $\zeta_{A1}=0.0001a_0^{-2}$ and $\beta=3.0$, and we use up to ten Gaussians of each angular-momentum type. To investigate the dependence of $\epsilon_b$ on the size of the positron basis set, we started with just a single $s$ function on each of the H, C, and N atoms, and then added further $s$ functions, one at a time, until the change in $\epsilon_b$ fell below 1\% (which required ten functions). A set of  $p$ functions with identical values of $\zeta_{Ak}$ was then added incrementally. Finally, a set of $d$ functions with identical values of $\zeta_{Ak}$ was added incrementally; only seven such  functions were required to achieve convergence.
\begin{table}
\caption{\label{tab:HCN_FT_RT_bin_en}Positron binding energy $\epsilon_b$ (a.u.) for HCN, in terms of the size of the positron basis set ($\zeta_{A1}=0.0001a_0^{-2}$ and $\beta=3.0$), in the FT and RT approximations. Negative values of $\epsilon_b$ indicate that the positron is not bound. Numbers in brackets indicate powers of 10.}
\begin{ruledtabular}
\begin{tabular}{lcc}
$e^+$ basis size & FT & RT \\
  \hline
 $1s$ & $-7.3104[-5]$ & $-7.2991[-5]$ \\
 $2s$ & $-5.0557[-5]$ & $-5.0399[-5]$ \\
 $3s$ & $-2.4132[-5]$ & $-2.3740[-5]$ \\
 $4s$ & $-7.2094[-6]$ & $-6.5130[-6]$ \\
 $5s$ & $1.3555[-5]$ & $1.4842[-5]$ \\
 $6s$ & $3.2674[-5]$ & $3.4758[-5]$ \\
  $7s$ & $5.2321[-5]$ & $5.5598[-5]$ \\
 $8s$ & $6.2984[-5]$ & $6.7133[-5]$ \\
$9s$ & $6.4272[-5]$ & $6.8536[-5]$ \\
 $10s$ & $6.4342[-5]$ & $6.8612[-5]$ \\
 $10s \, 1p$ & $6.7013[-5]$ & $7.1105[-5]$ \\
  $10s \, 2p$ & $6.7224[-5]$ & $7.1367[-5]$ \\
  $10s \, 3p$ & $6.7518[-5]$ & $7.1676[-5]$ \\
$10s \, 4p$ & $6.7741[-5]$ & $7.1931[-5]$ \\  
$10s \, 5p$ & $6.8100[-5]$ & $7.2332[-5]$ \\
$10s \, 6p$ & $6.8393[-5]$ & $7.2666[-5]$ \\
$10s \, 7p$ & $6.8516[-5]$ & $7.2806[-5]$ \\
$10s \, 8p$ & $6.8822[-5]$ & $7.3141[-5]$ \\
$10s \, 9p$ & $6.9130[-5]$ & $7.3479[-5]$ \\
$10s \, 10p$ & $6.9134[-5]$ & $7.3484[-5]$ \\
$10s \, 10p \, 1d$ & $7.0910[-5]$ & $7.4968[-5]$  \\
$10s \, 10p \, 2d$ & $7.1296[-5]$ & $7.5293[-5]$  \\
$10s \, 10p \, 3d$ & $7.1382[-5]$ & $7.5363[-5]$ \\
$10s \, 10p \, 4d$ & $7.1395[-5]$ & $7.5372[-5]$  \\
$10s \, 10p \, 5d$ & $7.1401[-5]$ & $7.5378[-5]$ \\
$10s \, 10p \, 6d$ & $7.1409[-5]$ & $7.5386[-5]$ \\
$10s \, 10p \, 7d$ & $7.1411[-5]$ & $7.5388[-5]$
\end{tabular}
\end{ruledtabular}
\end{table}

As expected, the RT value of $\epsilon_b$ is always greater than the FT value. However, the difference between them is very small, only 5\%, which shows that the weakly bound positron almost does not perturb the electron cloud. Our final RT value of $\epsilon_b=7.5388\times10^{-5}$~a.u. is in good agreement with the previous RT calculations of Chojnacki and Strasburger\cite{Chojnacki06} and Kita \textit{et al.},\cite{Kita09} which gave values of $6.0\times 10^{-5}$~a.u. and $7.3\times 10^{-5}$~a.u., respectively. The differences are due to using different values of the H--C and C${\equiv}$N bond lengths (1.066 and 1.167~\AA , respectively), and different electron and positron basis sets. 

Comparing the final $10s\,10p\,7d$ binding energy of $7.1411\times 10^{-5}$~a.u. (FT) or $7.5388\times 10^{-5}$~a.u. (RT) with the  $10s$ binding energy of $6.4342\times 10^{-5}$~a.u. (FT) or $6.8612\times 10^{-5}$~a.u. (RT), we observe that in spite of the large asymmetry of the dipole-bound state, the $s$ functions alone account for 90\% of the total binding energy. The $p$ functions provide about 7\% of $\epsilon_b$, while  the $d$ functions add 3\%. This is a result of placing positron basis functions on more than one center: linear combinations of $s$-type functions on multiple centers effectively generate higher-angular-momentum-type functions (see Appendix~\ref{sec:app_ang_mom}).\cite{Whitten63,*Whitten66,*Petke69} Thus, the basis set is already relatively complete before the true $p$- and $d$-type functions are added.

\begin{figure}
\centering
\includegraphics{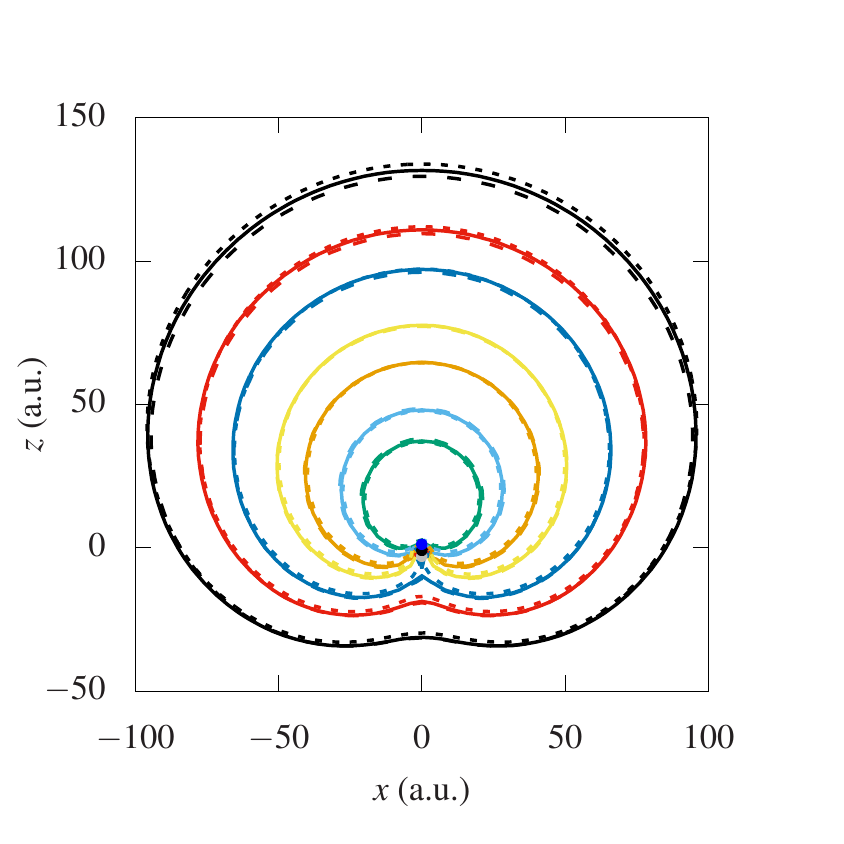}
\caption{\label{fig:FT_vs_RT_vs_MOD}Positron wave function $\psi(\vec{r})$ for $y=0$ in the FT and RT approximations and dipole model. The H, C, and N atoms are placed along the $z$ axis with coordinates $-2.921$, $-0.920$, and $1.209$~a.u., respectively. Solid contours, FT approximation; long-dashed contours, RT approximation; short-dashed contours, dipole model with $\mu=3.27$~D and $\epsilon_b=7.1411\times 10^{-5}$~a.u.\cite{Gribakin15}  The value of $\psi(\vec{r})$ on each contour is as follows (in a.u.): black, 0.0002; red, 0.0003; dark blue, 0.0004; yellow, 0.0006; orange, 0.0008; light blue, 0.0012; green, 0.0016.}
\end{figure}

The HCN molecule has $C_{\infty v}$ symmetry, so the positron wave function is symmetric with respect to rotation about the molecular axis $z$.
Figure~\ref{fig:FT_vs_RT_vs_MOD} shows the  $10s\,10p\,7d$ positron wave function $\psi(\vec{r})$ as a function of $x$ and $z$, with $y=0$, as calculated in the FT and RT approximations. The H, C, and N atoms are on the $z$ axis with coordinates $-2.921$, $-0.920$, and $1.209$~a.u., respectively.
The FT and RT wave functions are barely distinguishable on the scale of the graph. We see that the positron is strongly localized at the nitrogen end of the molecule, since this is the negatively charged end of the molecular dipole.

Figure~\ref{fig:FT_vs_RT_vs_MOD} also shows the positron wave function from the semianalytical ``dipole model'' developed to analyze positron binding to strongly polar molecules.\cite{Gribakin15} This model treats a polar molecule  as a point dipole with dipole moment $\vec{\mu}$, surrounded by an impenetrable sphere of radius $r_0$. The point dipole provides the long-range $\vec{\mu} \cdot \vec{r}/r^3$ potential for the positron, while the hard sphere mimics short-range repulsion by the atomic nuclei. The positron binding energy $\epsilon_b$ is in one-to-one correspondence with the sphere radius $r_0$, i.e., knowledge of the value of $\epsilon_b$ can be used to obtain the value of $r_0$, or vice versa. Note that this model does not use any information about the true geometry of the molecule.
Using $\mu=3.27$~D (the dipole moment of HCN at the HF level\footnote{The dipole moment obtained in Ref.~\onlinecite{Chojnacki06} is 3.312~D.}) and $\epsilon_b=7.1411\times 10^{-5}$~a.u. (the FT value), we find $r_0=1.98$~a.u. The resulting wave function, shown by a short-dashed curve in Fig.~\ref{fig:FT_vs_RT_vs_MOD}, is very close to FT and RT wave functions. This indicates that positron binding to a polar molecule at the static level is described well by a simple model of a point dipole enclosed by a hard sphere.

\subsection{FT+P approximation}\label{subsec:FTP}

For the FT+P calculations, we use the  atomic hybrid polarizabilities of Miller.\cite{Miller90} The values are $\alpha_\text{H}=0.387$~\AA$^3$, $\alpha_\text{C}=1.283$~\AA$^3$, and $\alpha_\text{N}=0.956$~\AA$^3$. This gives a total molecular polarizability of $2.63$~\AA$^3$, in near-exact agreement with the recommended value of $2.59$~\AA$^3$.\cite{CRC} For simplicity, we have chosen to take equal cutoff radii $\rho_A$ for the H, C, and N atoms. The choice of $\rho_A$ may look arbitrary at this stage, but values in the range 1.5--3.0~a.u. would be considered physical.\cite{Mitroy02}

Table~\ref{tab:FT+P_binding_energies} shows the binding energies obtained for $\rho_A=2.25$, 2.0, and 1.75~a.u.,
with smaller cutoff radii meaning a stronger correlation potential. The same parameters for the positron basis set have been used as in the FT and RT calculations.
\begin{table}
\caption{\label{tab:FT+P_binding_energies}Positron binding energy $\epsilon_b$ (a.u.) for HCN, in terms of the size of the positron basis set ($\zeta_{A1}=0.0001a_0^{-2}$ and $\beta=3.0$), in the FT+P approximation. Numbers in brackets indicate powers of 10.}
\begin{ruledtabular}
\begin{tabular}{lccc}
$e^+$ basis size & $\rho_A=2.25$~a.u. & $\rho_A=2.0$~a.u. & $\rho_A=1.75$~a.u. \\
\hline
$10s$                   & $1.1063[-3]$ & $1.6708[-3]$ & $2.9320[-3]$ \\
$10s \, 10p$         & $1.1426[-3]$ & $1.7205[-3]$ & $2.9988[-3]$ \\
$10s \, 10p \, 7d$ &  $1.1438[-3]$ & $1.7221[-3]$ & $2.9995[-3]$
\end{tabular}
\end{ruledtabular}
\end{table}
One can see that the final ($10s \, 10p \, 7d$) binding energy has increased by a factor of 16, 24, and 42, with respect to the static-dipole FT calculation, 
for $\rho_A=2.25$, 2.0, and 1.75~a.u., respectively. One can also see that including $p$- and $d$-type Gaussians has a smaller effect than in the static-dipole calculation. This is related to the fact that the wave function calculated with $V_\text{cor}$ becomes more spherical (see below).

The existing CI\cite{Chojnacki06} and diffusion Monte Carlo (DMC)\cite{Kita09} calculations gave $\epsilon_b=35$ and $38$~meV, respectively. These are closest to the binding energy of $1.1438\times 10^{-3}~\text{a.u.}\approx 31$~meV we obtained for $\rho_A=2.25$~a.u. However, as  CI and DMC  are variational methods, their predictions should be considered as lower bounds on the true binding energy. Thus, we believe that our result of $\epsilon_b=47$~meV obtained using $\rho_A=2.0$~a.u. (cf.
$\rho =2.05$~a.u. for atomic hydrogen\cite{Mitroy02}) may be closer to the true value of the positron binding energy for HCN. 

\begin{figure}
\centering
\includegraphics{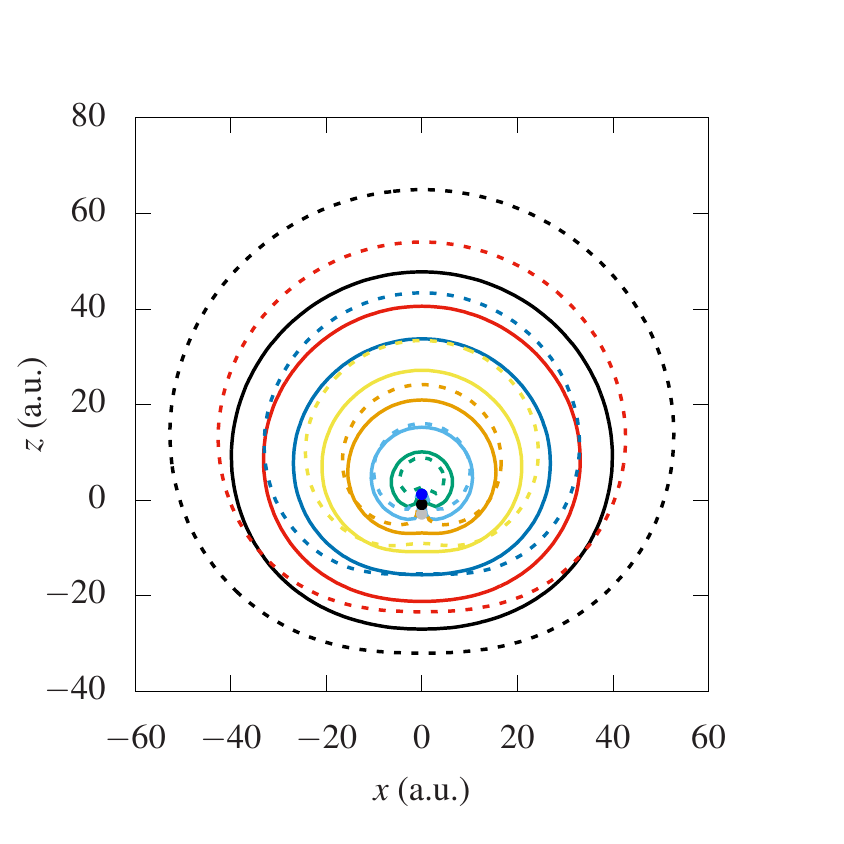}
\caption{\label{fig:1.75_vs_2.25}Positron wave function $\psi(\vec{r})$ for $y=0$ in the FT+P approximation. Dashed contours, $\rho_A=2.25$~a.u.; solid contours, $\rho_A=1.75$~a.u. The value of $\psi(\vec{r})$ on each contour is as follows (in a.u.): black, 0.0002; red, 0.0004; dark blue, 0.0008; yellow, 0.0016; orange, 0.0032; light blue, 0.0064; green, 0.0128.}
\end{figure}

Figure~\ref{fig:1.75_vs_2.25} shows  the $10s\,10p\,7d$ positron wave function $\psi(\vec{r})$ as a function of $x$ and $z$, with $y=0$, for $\rho_A=1.75$ and 2.25~a.u.
Comparing the scales on the axes of Fig.~\ref{fig:1.75_vs_2.25} and Fig.~\ref{fig:FT_vs_RT_vs_MOD}, we see that due to the effect of $V_\text{cor}$ and increased binding energy, the positron is found much closer to the molecule than in the static dipole approximation (FT or RT). This can also be seen from the position of the classical turning point on the positive $z$ axis in the dipole potential, $\mu /r^2=\epsilon _b$, which gives $r=134$~a.u. for the FT calculation, versus $r=33$~a.u. for the FT+P calculation with $\rho_A=2.25$~a.u. It is also evident that the wave function for $\rho_A=1.75$~a.u. ($\epsilon _b=82$~meV) is more compact compared with that for $\rho_A=2.25$~a.u. ($\epsilon _b=31$~meV).

To understand the shape of the wave function, consider a weakly bound  state in a short-range potential, such as $V_\text{cor}$ alone. The wave function away from the target would be spherically symmetric, $\psi (\vec{r})\sim \sqrt{\kappa /2\pi }e^{-\kappa r}/r$, where $\kappa =\sqrt{2\epsilon_b}$. 
In the FT approximation, the long-range dipole potential $V_\text{st}$ makes the positron wave function strongly asymmetric in the $z$ direction (Fig.~\ref{fig:FT_vs_RT_vs_MOD}). The addition of $V_\text{cor}$ in the FT+P approximation increases the binding energy significantly, making the long-range effect of $V_\text{st}$ less pronounced.
What we see in Fig.~\ref{fig:1.75_vs_2.25} in comparison with Fig.~\ref{fig:FT_vs_RT_vs_MOD} is a transition from a strongly asymmetric (in the $z$ direction) dipole-bound state, to a more spherically symmetric bound state that one would have had for a nonpolar molecule. However, in both calculations, the positron is strongly localized about the negatively charged nitrogen end of the molecule, despite the attraction to the H and C atoms provided by the correlation potential in FT+P. 

\begin{figure*}
\centering
\includegraphics[width=0.45\textwidth]{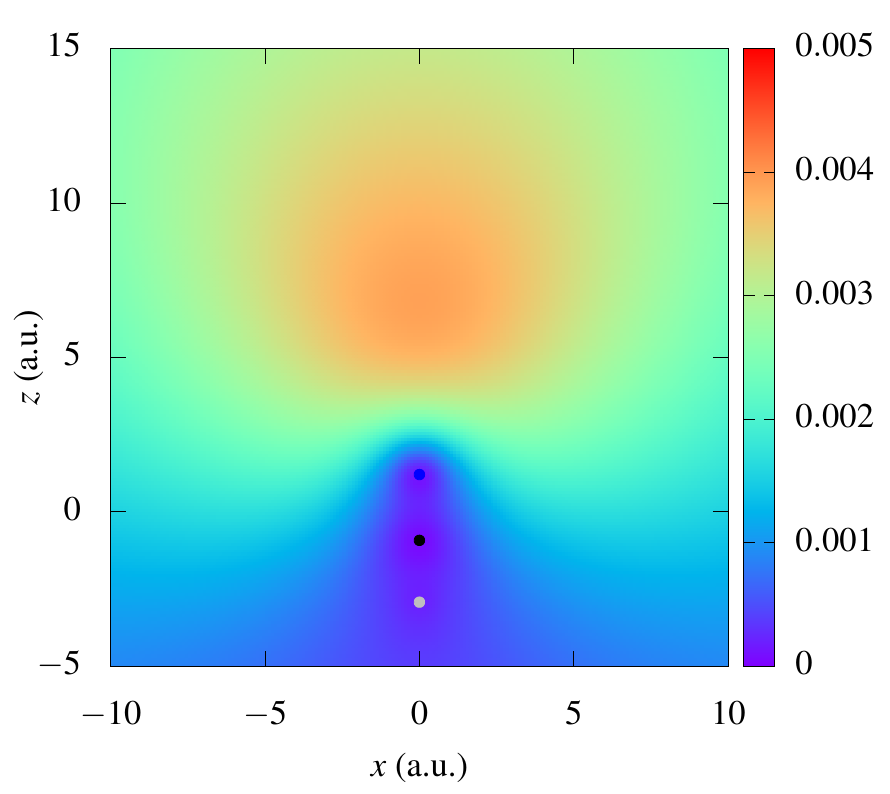}
\includegraphics[width=0.45\textwidth]{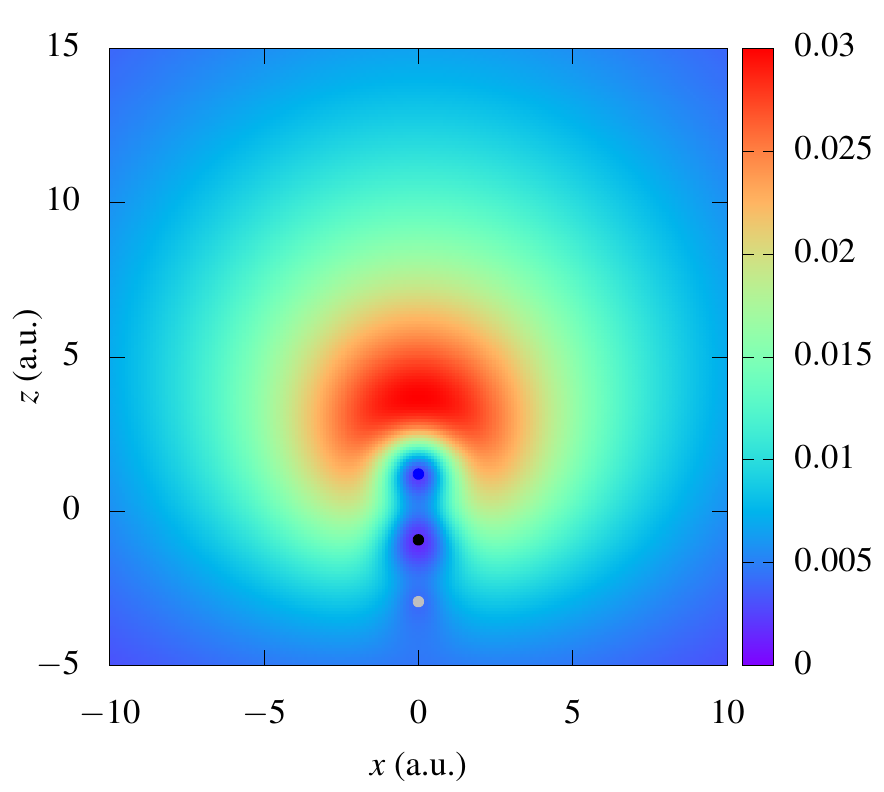}
\caption{\label{fig:FT_vs_FT+P_shortrange}Short-range behavior of the positron wave function $\psi(\vec{r})$ for $y=0$ in the FT (left) and FT+P ($\rho_A=1.75$~a.u., right) approximations.}
\end{figure*}

Going back to Table~\ref{tab:FT+P_binding_energies}, we notice that for all three values of $\rho_A$, the $s$-type basis functions alone contribute 97--98\% of the total binding energy. The $p$ functions contribute almost all of the remaining 2--3\%, with the contribution from the $d$ functions being essentially negligible. The inclusion of $p$ and $d$ functions is thus even less important in the FT+P calculation than it is in the FT or RT approximations. A possible explanation for this observation is as follows.  The positron wave function in the FT or RT calculation is strongly localized outside the nitrogen end of the molecule at both long and short range. This is also true for the long-range part of the FT+P wave function. However, at short range the FT+P wave function is more evenly spread over the whole molecule and ``more round'' near each of the atoms. This can be seen from Fig.~\ref{fig:FT_vs_FT+P_shortrange}, which compares the FT wave function with the FT+P wave function for $\rho_A=1.75$~a.u.
Consequently, a more significant proportion of the wave function is constructed from tight (i.e., large-exponent) $s$-type Gaussians in the FT+P approximation than in the FT approximation, with $p$ and $d$ functions playing a relatively minor role.

\begin{figure}
\centering
\includegraphics{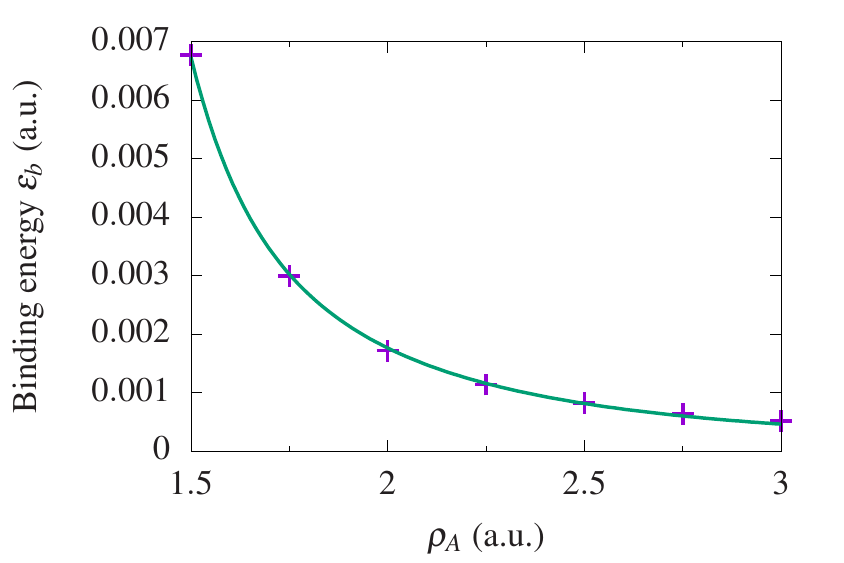}
\caption{\label{fig:eb_vs_rho}Positron binding energy as a function of the cutoff parameter $\rho_A$. Purple plusses, calculated values; green curve, empirical fit [Eq.~(\ref{eq:eb_vs_rho})].}
\end{figure}

Given the importance of the cutoff radius for the binding energy, we examine the dependence of $\epsilon_b$ on $\rho_A$ more closely in Fig.~\ref{fig:eb_vs_rho}. It shows $\epsilon_b$ for $\rho_A$ between 1.5 and 3.0~a.u., calculated using the $10s\,10p\,7d$ basis.
Also shown is the empirical fit
\begin{equation}\label{eq:eb_vs_rho}
\epsilon_b = \epsilon_b^\text{FT} + \frac{0.229}{\rho_A^{11.4}} + \frac{0.0179}{\rho_A^{3.48}} ,
\end{equation}
where $\epsilon_b^\text{FT}=7.1411\times10^{-5}$~a.u. is the FT value of $\epsilon_b$, which $\epsilon_b$ approaches in the limit $\rho_A \to \infty$. This fit is valid for $\rho_A\geq 1.5$~a.u.; applying Eq.~(\ref{eq:eb_vs_rho}) for $\rho_A< 1$~a.u. would yield unphysically large values of $\epsilon_b$.
 
Figure~\ref{fig:eb_vs_rho} shows that the binding energy is sensitive to the choice of $\rho_A$. Using values in the physically plausible range $1.75\leq \rho_A\leq 2.25$~a.u., results in a factor of two uncertainty of the binding energy, which seems quite acceptable for a model-potential theory.

\begin{figure}
\centering
\includegraphics{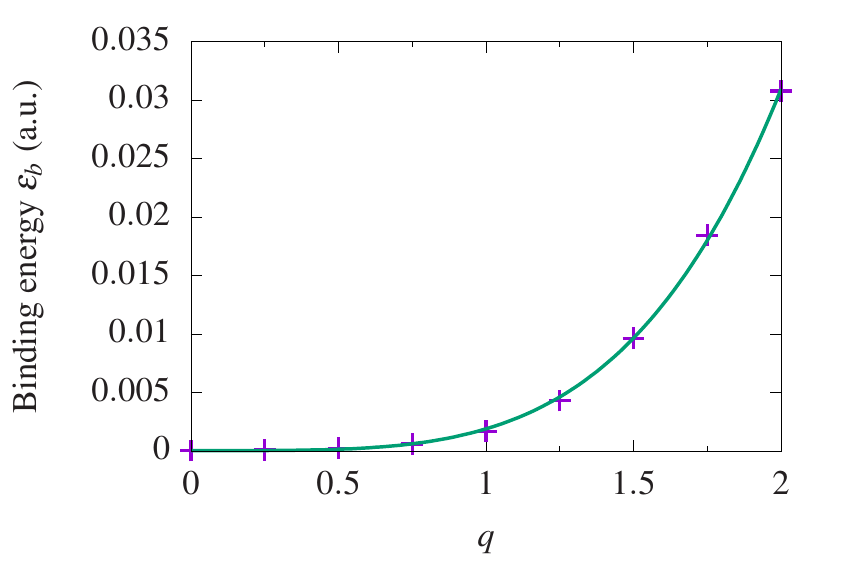}
\caption{\label{fig:eb_vs_alpha}Positron binding energy as a function of the polarizability scaling factor $q$, for $\rho_A=2.0$~a.u. Purple plusses, calculated values; green curve, empirical fit [Eq.~(\ref{eq:eb_vs_alpha})].}
\end{figure}

It is also useful to investigate the sensitivity of the binding energy to the value of the molecular polarizability, for a fixed value of the cutoff parameter. We do this by multiplying $V_\text{cor}$ used in the calculations by a dimensionless factor $q$. Figure~\ref{fig:eb_vs_alpha} shows $\epsilon_b$ for $q$ between 0 and 2, for a fixed value of $\rho_A=2.0$~a.u.
It also shows an empirical power-law fit,
\begin{equation}\label{eq:eb_vs_alpha}
\epsilon_b \approx 0.00185 q^{4.06},
\end{equation}
valid away from the origin. Equation~(\ref{eq:eb_vs_alpha}) shows that a 5\% uncertainty in the value of the molecular dipole polarizability (or the magnitude of $V_\text{cor}$) would result in a 20\% uncertainty of the positron-molecule binding energy.

The above analysis quantifies the strong sensitivity of the positron-molecule binding energies to the magnitude of the correlation potential, i.e., to the extent that electron-positron correlations are included in the calculation. This highlights the difficulty faced by \textit{ab initio} approaches in predicting positron-molecule binding energies. On the other hand, we see that our model accurately captures the essential physics of the bound positron-molecule system. Using physically acceptable values of the dipole polarizability and cutoff parameter, we obtain values of $\epsilon_b$ in good agreement with existing state-of-the-art calculations that account for dynamic electron-positron correlations.

\subsection{Annihilation rate}\label{subsec:ann_rate}

The wave functions of the positron bound state obtained in the FT and FT+P calculations can be used to estimate the electron-positron contact density $\delta_{ep}$ using Eq.~(\ref{eq:con_den_IPA}). In the FT+P case, we also account for the short-range electron-positron correlations that increase the electron density at the positron, by using Eq.~(\ref{eq:con_den_enh}) together with Eq.~(\ref{eq:enh_fac_formula}) for the enhancement factors. 

Table~\ref{tab:contact_density} shows the contact densities obtained  in both the FT and FT+P approximations in terms of the size of the positron basis set.  For the FT+P calculations,  values of $\rho_A=2.25$, 2.0, and 1.75~a.u. were used, and both the unenhanced and enhanced results are shown. 
\begin{table*}[ht!!]
\caption{\label{tab:contact_density}Electron-positron contact density $\delta_{ep}$ (a.u.) for HCN, in terms of the size of the positron basis set ($\zeta_{A1}=0.0001a_0^{-2}$ and $\beta=3.0$), in the FT and FT+P approximations. For the FT+P approximation, both the unenhanced [Eq.~(\ref{eq:con_den_IPA})] and enhanced [Eq.~(\ref{eq:con_den_enh})] values are shown. Numbers in brackets indicate powers of 10.
}
\begin{ruledtabular}
\begin{tabular}{lccccccc}
&& \multicolumn{3}{c}{FT+P (unenhanced)} & \multicolumn{3}{c}{FT+P (enhanced)} \\
\cline{3-5} \cline{6-8}
$e^+$ basis size & FT & $\rho_A=2.25$~a.u. & $\rho_A=2.0$~a.u. & $\rho_A=1.75$~a.u. & $\rho_A=2.25$~a.u. & $\rho_A=2.0$~a.u. & $\rho_A=1.75$~a.u. \\
\hline
$10s$               & $1.0313[-5]$ & $4.8902[-4]$ & $8.7874[-4]$ & $1.8844[-3]$ & $2.2500[-3]$ & $4.0216[-3]$ & $8.5455[-3]$    \\
$10s\,10p$       & $1.0172[-5]$ & $4.9673[-4]$ & $8.9122[-4]$ & $1.9030[-3]$ & $2.2820[-3]$ & $4.0722[-3]$ & $8.6173[-3]$   \\
$10s\,10p\,7d$ & $9.6738[-6]$ & $4.9718[-4]$ & $8.9171[-4]$ & $1.9030[-3]$ & $2.2846[-3]$ & $4.0753[-3]$ & $8.6178[-3]$ 
\end{tabular}
\end{ruledtabular}
\end{table*}
We observe that the inclusion of $V_\text{cor}$ in the FT+P calculations increases
the contact densities by two orders of magnitude, compared with static-dipole FT values (even before the enhancement factors are used). This is a direct result of the significantly stronger binding in the FT+P approximation: the attractive correlation potential draws the positron wave function in (see Figs.~\ref{fig:FT_vs_RT_vs_MOD} and \ref{fig:1.75_vs_2.25}), greatly increasing the positron density near the molecule. In turn, including the enhancement factors produces $\delta_{ep}$ values that are about a factor of 4.5 greater than their unenhanced counterparts.

As the size of the positron basis set increases, the FT+P contact densities all increase. This is as expected: increasing the completeness of the basis results in stronger binding, and therefore, greater positron density near the molecule. The $s$ functions alone provide 98--99\% of the final $10s\,10p\,7d$ value of the contact density. The $p$ functions provide almost all of the remaining 1--2\%, while the $d$ functions have a negligible contribution.

The FT contact densities display the opposite trend: the contact density actually \textit{decreases} as the size of the positron basis increases. Moreover, while the contact density in the $10s\,10p$ calculation is merely 1\% smaller than the $10s$ value, the $10s\,10p\,7d$ value is some 6\% smaller than the $10s$ value. 
To understand this, recall that in the FT approximation, including the $p$- and $d$-type Gaussians contributes much more significantly to the binding energy than in the FT+P calculation (see Tables~\ref{tab:HCN_FT_RT_bin_en} and \ref{tab:FT+P_binding_energies}). It follows that the FT wave function has a greater contribution from $p$ and $d$ Gaussians, compared with the FT+P wave function. As mentioned in Sec.~\ref{sec:FT_RT_results}, the long-range behavior of the diffuse wave function of the dipole-bound state is described well by the $s$-type Gaussians placed on the three centers.
However, $s$-type Gaussians take finite values at their origins, i.e., at the positions of the atoms. The role of the $p$ and $d$ functions is thus to ``take over'' the description of the long-range behavior from the $s$ functions, and ensure that the wave function is described correctly at short range, where it is strongly affected by the repulsion from the atomic nuclei.

Our best prediction of the annihilation rate (\ref{eq:Gamma_from_delta}) in the bound state is obtained using the FT+P enhanced contact density for $\rho_A=2.25$~a.u. ($\epsilon _b=31$~meV) and 2.00~a.u. ($\epsilon _b=47$~meV), which gave the binding energies in closest agreement with the existing calculations. Using the $10s\,10p\,7d$ values of $\delta_{ep}$ from Table~\ref{tab:contact_density}, we predict 
$\Gamma = 0.115 \times10^9~\text{s}^{-1}$ for $\epsilon _b=31$~meV, and 
$\Gamma = 0.206\times10^9~\text{s}^{-1}$ for $\epsilon _b=47$~meV.


\section{Conclusions}

Calculation of positron binding to polyatomic molecules is a difficult problem because of the extreme importance of electron-positron correlations. Solving this problem accurately appears to be beyond the capability of standard quantum chemistry approaches. As a result, a large body of experimental data on positron binding and annihilation in polyatomic molecules remains largely unexplained. In particular, trends in positron binding energies across various molecular families and the origin
of empirical relation between the binding energy and molecular parameters, such as the dipole polarizability and dipole moment, are poorly understood.

In this paper we have developed an approach that allows calculations of positron binding to both polar and nonpolar molecular species. Its key element is inclusion of a physically motivated model correlation potential that acts on the positron and accounts for the long-range polarization and short-range correlations. The potential contains short-range cutoff parameters that can be viewed as free parameters of the theory. However, their values are strongly constrained by accurate calculations of positron scattering and binding with atoms.

As a first application, positron binding to the HCN molecule has been explored. Being a strongly polar molecule, HCN binds the positron even at the level of a static-potential approximation, with a binding energy of about 2~meV. Our calculations showed that positron binding in the static-dipole approximation is described very well by a simple model,\cite{Gribakin15} in which the molecule is replaced by a point dipole surrounded by a hard sphere. Adding the correlation potential confirmed strong enhancement of binding due to correlation effects, seen earlier in the CI\cite{Chojnacki06} and QMC\cite{Kita09} calculations. Moreover, a physically motivated choice of the cutoff parameter yielded the binding energy in good accord with the above calculations. We also used the wave function of the bound state to calculate the annihilation rate, including important short-range correlation enhancement factors.\cite{Green15,Green18}

Although our description of the bound positron-molecule system is not \textit{ab initio}, its simplicity enables clear physical insight into the problem. The model correlation potential contains at most one free parameter for each type of atom in the molecule: the cutoff radius (assuming that the values of the hybrid polarizabilities of the atoms are known). The real aim of our approach is to explore positron binding to larger polyatomic molecules, in particular, to nonpolar species for which presently there are no calculations. We plan to use a small subset of experimentally known binding energies to ``calibrate'' our correlation potential, i.e., determine 
the cutoff radius for the C and H atoms, which would enable calculations for various alkane molecules.

Calculations can then be extended to alkane rotamers, aromatic hydrocarbons, and other hydrocarbons that support binding (e.g, ethylene and acetylene). Bringing into consideration the cutoff radius for an O atom will enable calculations for alcohols, aldehydes, ketones, formates, and acetates. Likewise, considering the N atom will enable a study of the nitriles. Thus, it is hoped that accurate calculations of the positron binding energy will be possible for the vast majority of the molecules for which they have been measured. In addition to the annihilation rates, we will also use the bound-state positron wave functions to compute annihilation $\gamma$-ray spectra, where much of the experimental data\cite{Iwata97} remained unexplained for a long time\cite{Green12} and have only started to be explored now.\cite{Ikabata18}

\begin{acknowledgments}
This work has been supported by the EPSRC UK, Grant No. EP/R006431/1.
\end{acknowledgments}

\appendix
\section{\label{sec:app_a}Calculation of matrix elements of correlation potential and electron-positron contact density}

Using Eqs.~(\ref{eq:gaussian_def}), (\ref{eq:pol_breakdown}), and (\ref{eq:pol_pot_partial_expansion}), along with the Gaussian product rule, 
\begin{align}\label{eq:gaussian_product_rule}
e^{-\zeta_1 \lvert \vec{r} - \vec{r}_1\rvert^2}e^{-\zeta_2 \lvert \vec{r} - \vec{r}_2\rvert^2} &= \exp\left(-\frac{\zeta_1 \zeta_2}{\zeta_1 + \zeta_2}\lvert \vec{r}_1-\vec{r}_2\rvert^2 \right) \nonumber \\
& \quad{}\times \exp\left[-(\zeta_1 + \zeta_2) \left\lvert \vec{r}-\frac{\zeta_1\vec{r}_1 + \zeta_2 \vec{r}_2}{\zeta_1 + \zeta_2} \right\rvert^2 \right] ,
\end{align}
a matrix element of $V_\text{cor}$ between positron basis functions $g_{Ak}$ and $g_{Bl}$ is given by
\begin{align}
\langle g_{Bl} \vert V_\text{cor} \vert g_{Ak} \rangle &= \sum_{C=1}^{N_a} \langle g_{Bl} \vert V_\text{cor}^{(C)} \vert g_{Ak} \rangle \nonumber\\
&=\sum_{C=1}^{N_a} \sum_m D_{m}^{(C)}\!\int \!g_{Bl}^*(\vec{r}) e^{-\kappa_{Cm} \lvert \vec{r}-\vec{r}_C\rvert^2} g_{Ak}(\vec{r})\, d^3\vec{r} \nonumber\\
&= F^*_{Bl} F_{Ak} e^{-\nu \lvert \vec{r}_A-\vec{r}_B\rvert^2} \nonumber\\
&\quad{}\times \sum_{C=1}^{N_a} \sum_m D_{m}^{(C)} e^{-\lambda \lvert \vec{r}_C - \vec{r}_{AB}\rvert^2}
H_x H_y H_z,
\end{align}
where 
\begin{align}
\nu &= \frac{\zeta_{Ak} \zeta_{Bl}}{\zeta_{Ak} +\zeta_{Bl}} , \\
\lambda &= \frac{(\zeta_{Ak} +\zeta_{Bl})\kappa_{Cm}}{\zeta_{Ak} +\zeta_{Bl}+\kappa_{Cm}} , \\
\vec{r}_{AB} &= \frac{\zeta_{Ak}\vec{r}_A + \zeta_{Bl}\vec{r}_B}{\zeta_{Ak} +\zeta_{Bl}} , \\
\vec{r}_{ABC} &= \frac{\zeta_{Ak}\vec{r}_A +\zeta_{Bl}\vec{r}_B+\kappa_{Cm}\vec{r}_C}{\zeta_{Ak} +\zeta_{Bl}+\kappa_{Cm}} ,\\
H_\mu &= \int_{-\infty}^\infty (\mu-\mu_A)^{n^\mu_{Ak}} (\mu-\mu_B)^{n^\mu_{Bl}} \nonumber\\
&\quad{}\times \exp[-(\zeta_{Ak}+\zeta_{Bl}+\kappa_{Cm} ) (\mu-\mu_{ABC})^2] \, d\mu .
\end{align}
The integral $H_\mu$ is evaluated analytically by ``translating'' the polynomials to position $\mu_{ABC}$, e.g.,
\begin{align}\label{eq:binomial_expansion}
(\mu-\mu_A)^{n^\mu_{Ak}} &= (\mu - \mu_{ABC} + \mu_{ABC} - \mu_A)^{n^\mu_{Ak}} \nonumber\\
&= \sum_{s^\mu_{Ak}=0}^{n^\mu_{Ak}}
\binom{n^\mu_{Ak}}{s^\mu_{Ak}}
(\mu-\mu_{ABC})^{s^\mu_{Ak}}
(\mu_{ABC}-\mu_A)^{n^\mu_{Ak}-s^\mu_{Ak}},
\end{align}
and using the identity
\begin{align}\label{eq:gamma_identity}
\int_{-\infty}^{\infty} \xi^n e^{-a \xi^2}  d\xi
= \frac12 [1+(-1)^n] a^{-(1+n)/2} \Gamma \left( \frac{1+n}{2} \right),
\end{align}
which is valid for $n=0,1,2,\dotsc$ and $a>0$. We obtain
\begin{align}
H_\mu &= \sum_{s^\mu_{Ak}=0}^{n^\mu_{Ak}}
\sum_{s^\mu_{Bl}=0}^{n^\mu_{Bl}}
\binom{n^\mu_{Ak}}{s^\mu_{Ak}}
\binom{n^\mu_{Bl}}{s^\mu_{Bl}}
\frac12 \big[1 + (-1)^{s^\mu_{Ak} + s^\mu_{Bl}} \big]
\nonumber\\
&\quad{}\times
(\mu_{ABC} - \mu_A)^{n^\mu_{Ak} - s^\mu_{Ak}}
(\mu_{ABC} - \mu_B)^{n^\mu_{Bl} - s^\mu_{Bl}} \nonumber\\
&\quad{}\times
(\zeta_{Ak}+\zeta_{Bl}+\kappa_{Cm} )^{-(1+s^\mu_{Ak} + s^\mu_{Bl})/2} \nonumber\\
&\quad{}\times
\Gamma \left(  \frac{1+s^\mu_{Ak} + s^\mu_{Bl}}{2}  \right) \qquad (\mu=x,y,z).
\end{align}

Using Eqs.~(\ref{eq:ele_bas_exp}), (\ref{eq:pos_bas_exp}), (\ref{eq:gaussian_def}), (\ref{eq:gaussian_product_rule}), (\ref{eq:binomial_expansion}), and (\ref{eq:gamma_identity}), the electron-positron contact density, Eq.~(\ref{eq:con_den_enh}), is given by
\begin{align}
\delta_{ep} &= 2 \sum_{i=1}^{N_e/2} \gamma_i
\sum_{A=1}^{N_a} \sum_{k=1}^{N^e_A}
\sum_{A'=1}^{N_a} \sum_{k'=1}^{N^e_{A'}}
\sum_{B=1}^{N_a} \sum_{l=1}^{N^p_B}
\sum_{B'=1}^{N_a} \sum_{l'=1}^{N^p_{B'}}
 \nonumber\\
&\quad{}\times
C_{Ak}^{(i)} C_{A'k'}^{(i)*}
C_{Bl}^{(p)} C_{B'l'}^{(p)*} 
F_{Ak}F_{A'k'}^* F_{Bl} F_{B'l'}^* \,e^{-\sigma_i\lvert \vec{r}_A-\vec{r}_{A'}\rvert^2}   \nonumber\\
&\quad{}\times
e^{-\sigma_p\lvert \vec{r}_B-\vec{r}_{B'}\rvert^2}
e^{-\tau \lvert \vec{r}_{AA'} - \vec{r}_{BB'}\rvert^2} I_x I_y I_z ,
\end{align}
where
\begin{align}
\sigma_i  &= \frac{\zeta_{Ak} \zeta_{A'k'}}{\zeta_{Ak} + \zeta_{A'k'}} , \\
\sigma_p &= \frac{\zeta_{Bl} \zeta_{B'l'}}{\zeta_{Bl} + \zeta_{B'l'}} , \\
\tau &= \frac{(\zeta_{Ak} + \zeta_{A'k'})(\zeta_{Bl} + \zeta_{B'l'})}{\zeta_{Ak} + \zeta_{A'k'}+\zeta_{Bl} + \zeta_{B'l'}} , \\
\vec{r}_{AA'} &= \frac{\zeta_{Ak} \vec{r}_A + \zeta_{A'k'} \vec{r}_{A'}}{\zeta_{Ak} + \zeta_{A'k'}} , \\
\vec{r}_{BB'} &= \frac{\zeta_{Bl} \vec{r}_B + \zeta_{B'l'} \vec{r}_{B'}}{\zeta_{Bl} + \zeta_{B'l'}} , \\
\vec{r}_{AA'BB'} &= \frac{\zeta_{Ak}\vec{r}_A + \zeta_{A'k'}\vec{r}_{A'}+\zeta_{Bl}\vec{r}_B + \zeta_{B'l'}\vec{r}_{B'}}{\zeta_{Ak} + \zeta_{A'k'}+\zeta_{Bl} + \zeta_{B'l'}} , 
\end{align}
and
\begin{widetext}
\begin{align}
I_\mu &= \int_{-\infty}^{\infty} (\mu - \mu_A)^{n^\mu_{Ak}} (\mu - \mu_{A'})^{n^\mu_{A'k'}}  (\mu - \mu_B)^{n^\mu_{Bl}} (\mu - \mu_{B'})^{n^\mu_{B'l'}}
 \exp[-(\zeta_{Ak} + \zeta_{A'k'}+\zeta_{Bl} + \zeta_{B'l'}) (\mu - \mu_{AA'BB'})^2] \, d\mu \nonumber\\
&= \sum_{s^\mu_{Ak}=0}^{n^\mu_{Ak}}
\sum_{s^\mu_{A'k'}=0}^{n^\mu_{A'k'}}
\sum_{s^\mu_{Bl}=0}^{n^\mu_{Bl}}
\sum_{s^\mu_{B'l'}=0}^{n^\mu_{B'l'}} 
\binom{n^\mu_{Ak}}{s^\mu_{Ak}}
\binom{n^\mu_{A'k'}}{s^\mu_{A'k'}}
\binom{n^\mu_{Bl}}{s^\mu_{Bl}}
\binom{n^\mu_{B'l'}}{s^\mu_{B'l'}}
\frac12 \big[1+(-1)^{ s^\mu_{Ak}+ s^\mu_{A'k'}+ s^\mu_{Bl}+ s^\mu_{B'l'}} \big]
\nonumber\\
&\quad{}\times(\mu_{AA'BB'} - \mu_A)^{n^\mu_{Ak} - s^\mu_{Ak}}
(\mu_{AA'BB'} - \mu_{A'})^{n^\mu_{A'k'} - s^\mu_{A'k'}}
(\mu_{AA'BB'} - \mu_B)^{n^\mu_{Bl} - s^\mu_{Bl}}
(\mu_{AA'BB'} - \mu_{B'})^{n^\mu_{B'l'} - s^\mu_{B'l'}} \nonumber\\
&\quad{}\times
(\zeta_{Ak} + \zeta_{A'k'}+\zeta_{Bl} + \zeta_{B'l'})^{-(1+ s^\mu_{Ak}+ s^\mu_{A'k'}+ s^\mu_{Bl}+ s^\mu_{B'l'})/2} \Gamma \left( \frac{1+ s^\mu_{Ak}+ s^\mu_{A'k'}+ s^\mu_{Bl}+ s^\mu_{B'l'}}{2}\right) \qquad (\mu=x,y,z).
\end{align}
\end{widetext}

\section{\label{sec:app_ang_mom}Generation of higher-angular-momentum-type Gaussians from multicenter \textit{s}-type Gaussians}

Consider two $s$-type Gaussians with a common exponent $\zeta$, placed on the $z$ axis at positions $\pm a$. One particular linear combination of these functions (ignoring normalization constants) is 
\begin{align}
f(\vec{r},a) &=  e^{-\zeta \lvert \vec{r} - \vec{a}\rvert^2} -  e^{-\zeta \lvert \vec{r} + \vec{a}\rvert^2} \nonumber\\
&= 2 e^{-\zeta a^2} e^{-\zeta r^2} \sinh 2\zeta a z ,
\end{align}
where  $\vec{a} = a\vec{\hat{k}}$ and $\vec{\hat{k}}$ is a unit vector in the positive $z$ direction. Expanding  to first order around $a=0$ gives
\begin{equation}\label{eq:two_s_expansion}
f(\vec{r},a) \simeq 4\zeta a   z e^{-\zeta r^2} ,
\end{equation}
which is an effective $p_z$-type Gaussian centered on the origin. 

Now consider the following linear combination of three $s$-type Gaussians, placed at $z=0$, $\pm a$:
\begin{align}
h(\vec{r},a) &= e^{-\zeta \lvert\vec{r}-\vec{a}\rvert^2} -2e^{-\zeta a^2} e^{-\zeta r^2} + e^{-\zeta \lvert\vec{r}+\vec{a}\rvert^2} \nonumber\\
&= 2 e^{-\zeta a^2} e^{-\zeta r^2} (\cosh 2\zeta a z - 1) .
\end{align}
Expanding to second order around $a=0$ gives
\begin{equation}\label{eq:three_s_expansion}
h(\vec{r},a) \simeq 4 \zeta^2 a^2  z^2 e^{-\zeta r^2},
\end{equation}
which is an effective $d_{zz}$-type Gaussian centered on the origin. 

Equations~(\ref{eq:two_s_expansion}) and (\ref{eq:three_s_expansion}) are valid provided $r \gg a$. It can similarly be shown that placing $l$ Gaussians of $s$ type with the same exponent $\zeta$ at equally spaced centers along the $z$ axis generates an effective $z^l \exp(-\zeta r^2)$ Gaussian   at the midpoint of the centers. To obtain effective Gaussians with a nonzero projection of angular momentum along the $z$ axis would require centers off the $z$ axis.\cite{Whitten63,Whitten66,Petke69}


%

\end{document}